\documentclass[]{aa}
\usepackage{txfonts,graphicx,natbib,subfigure}
\bibpunct{(}{)}{;}{a}{}{,} 
\begin{document}
\title{Chromospheric magnetic reconnection: Two-fluid simulations of coalescing current loops}
\author{P.~D. Smith\inst{1} \and J.~I. Sakai\inst{2}}
\titlerunning{Chromospheric magnetic reconnection}
\authorrunning{P. D. Smith \and J.~I. Sakai}
\offprints{\\ Phil Smith, \email{P.D.Smith@pgr.salford.ac.uk}}
\institute{Institute for Materials Research, University of Salford, Greater Manchester, M5 4WT, United Kingdom. \and Laboratory for Plasma Astrophysics, Faculty of Engineering, University of Toyama, 3190, Gofuku, Toyama, 930-8555, Japan.}
\date{Received Xerox / Accepted Xerox}

\abstract {}{To investigate magnetic reconnection rates during the coalescence of two current loops in the solar chromosphere, by altering the neutral-hydrogen to proton density ratio, ioniziation/recombination coefficients, collision frequency and relative helicity of the loops.}{2.5D numerical simulations of the chromosphere were conducted using a newly developed two-fluid (ion-neutral) numerical code. Developed from the Artificial Wind scheme, the numerical code includes the effects of ion-neutral collisions, ionization/recombination, thermal/resistive diffusivity and collisional/resistive heating.}{It was found that the rates of magnetic reconnection strongly depend on the neutral-hydrogen to proton density ratio; increasing the density ratio by a thousand-fold decreased the rate of magnetic reconnection by twenty-fold. This result implies that magnetic reconnection proceeds significantly faster in the upper chromosphere, where the density of ions (protons) and neutral-hydrogen is comparable, than in the lower chromosphere where the density of neutral-hydrogen is over a thousand times the ion density. The result also implies that jets associated with fast magnetic reconnection, occur preferentially in the upper chromosphere / lower corona. The inclusion of ionization/recombination, an important physics effect in the chromosphere, increases the total reconnected magnetic flux, but does not alter the rate of magnetic reconnection. Reductions in the ion-neutral collision frequency, result in small increases to the rates of magnetic reconnection. The relative helicity of the two current loops was not observed to have any significant effect on the rates of magnetic reconnection. Comparisons of two-fluid and MHD (MagnetoHydroDynamic) simulations, show significant differences in the measured rates of magnetic reconnection, particularly for the higher neutral density cases which represent the lower chromosphere. This demonstrates that MHD is not an appropriate model for simulating magnetic reconnection in the solar chromosphere.}{The magnetic reconnection rates of coalescing current loops are strongly affected by the inclusion of neutral-hydrogen particles. It is therefore essential that ion-neutral collisions are included in future analytical/numerical models of chromospheric magnetic reconnection.}


\maketitle

\section{Introduction}

Magnetic reconnection is an important physical process in almost all cosmological and laboratory plasmas, because it serves as a mechanism for converting stored magnetic field energy into heat and non-thermal plasma energy. Magnetic reconnection in fully ionized plasmas has been extensively investigated using the resistive-MHD \citep{Priest_Forbes_2000, Biskamp_2005} and the Hall-MHD approximations, as well as in collisionless plasmas. The study of collisionless magnetic reconnection is well summarized by both the GEM \citep{Birn_et_al_2001} and Newton \citep{Birn_et_al_2005} reconnection challenges. Where in each challenge, a variety of different numerical codes were used to simulate the coalescence of magnetic islands; a Harris-type equilibrium.

Magnetic reconnection in weakly ionized plasmas is important in the solar photosphere/chromosphere \citep{Sakai_1996,Bulanov_Sakai_1998,Litvinenko_1999,Furusawa_Sakai_2000,Sakai_et_al_2006}, as well as in the interstellar medium \citep{Dorman_Kulsrud_1995,Zweibel_Brandenburg_1997,Heitsch_Zweibel_2003}. Solar observations of spectroscopic line shifts and non-thermal broadening, confirm that the photosphere, chromosphere and lower corona of the Sun are all in a highly dynamical state. These indicate that mass supply and wave propagation are both important mechanisms for transferring energy from the photosphere/chromosphere to the corona and ultimately the solar wind. Studies conducted by SUMER (Solar Ultraviolet Measurements of Emitted Radiation; \citet{Wilhelm_et_al_1995}), part of the SOHO mission, clarified that observed explosive events are bi-directional jets generated by magnetic reconnection \citep{Innes_et_al_1997,Innes_Toth_1999,Roussev_Galsgaard_2002}. Observations have also shown evidence of magnetic reconnection in the chromosphere and photosphere \citep{Bubio_Beck_2005,McIntosh_2007}, where the neutral-hydrogen to proton density ratio changes from one to $10^{4}$. Recent observations by \citet{Shibata_et_al_2007} using the SOT (Solar Optical Telescope) instrument onboard the Hinode satellite, show ubiquitous anemone jets in the chromospheres of active regions; strong indirect evidence for the existence of chromospheric magnetic reconnection.

Previously, in an attempt to better understand the observed transient phenomena in the solar chromosphere, \citet{Sakai_et_al_2006} developed a simulation code to describe the dynamics of two-fluid (ion-neutral) plasmas. The two fluids, are coupled through proton/neutral-hydrogen collisions, as well as through ionization/recombination. \citet{Sakai_et_al_2006} were successfully able to simulate the coalescence of of two counter-helical current loops (magnetic flux tubes) in the upper chromosphere. They found that the dynamics of two-fluid plasmas are quite different from those described by the single-fluid MHD approximation. During the simulated coalescence of the two current loops, they also observed a number of dynamical effects, including; proton heating, the formation of bi-directional proton jets and slow bi-directional plasma flows.

In this study, we will attempt to extend the \citet{Sakai_et_al_2006} investigation into the magnetic reconnection of coalescing current loops. To achieve this, we have developed a new two-fluid numerical code, which we use to investigate how the magnetic reconnection rates of the coalescing current loops is altered by changing the; (a) neutral-hydrogen to proton density ratio, (b) ionization/recombination coefficients, (c) ion-neutral collision frequency, and (d) relative helicities of the two current loops. The results will be applicable to both chromospheric and photospheric magnetic reconnection studies.

In Sect.~2 we present the two-fluid equations used to describe the dynamics of ion and neutral particles. In Sect.~3 we describe the numerical scheme used in our newly developed two-fluid code. In Sect.~4 we describe the simulation model and initial conditions used in this study. In Sect.~5 we present our simulation results and finally in Sect.~6 we discuss the conclusions that can be drawn from this study.

\section{Two-fluid Equations}

In this section we present the basic equations used to describe two-fluid (ion-neutral) plasmas. In what follows, the plasma density, pressure, velocity and magnetic field are given respectively by $\rho$, $P$, $\vec{V}$ and $\vec{B}$, where the subscripts \emph{p} and \emph{n} refer to the ion (proton) and neutral fluids. We begin with the equations for the neutral fluid
\begin{eqnarray}
\frac{\partial \rho_n}{\partial t} + \nabla \cdot \left( \rho _n \vec{ V}_n \right) & = & - S_{1}, \\
\frac{\partial \left(\rho_n \vec{ V}_n \right) }{\partial t} + \nabla \cdot \left( \rho_n \vec{ V}_n \vec{ V}_n \right) + \nabla P_n  & = & - S_{2},\\
\frac{\partial \varepsilon_n }{\partial t} + \nabla \cdot \left[ {\left( {\varepsilon _n + P_n } \right)\vec{ V}_n } \right] - q_{n} & = & - S_{3},
\label{eqn_123}
\end{eqnarray}
where the neutral plasma energy is given by
\begin{equation}
\varepsilon_n = \frac{P_n}{\gamma - 1} + \frac{\rho_n |\vec{V}_n|^2}{2},
\label{eqn_4}
\end{equation}
and the adiabatic constant is $\gamma = 5/3$. The coefficients $\alpha_i$, $\alpha_r$ and $\alpha_c$ refer respectively to the effects of ionization, recombination and ion-neutral collisions. The neutral heat flux is given by $q_n = \nabla ^2 \left( \lambda P_n / \rho _n \right)$, where $\lambda$ is the heat transfer constant. The source terms $S_{1}$, $S_{2}$ and $S_{3}$ seen above, collectively describe the effects of ionization/recombination, ion-neutral drag and collisional heating
\begin{eqnarray}
S_{1} & = & - \rho_p \left( \alpha_r \rho_p  - \alpha_i \rho_n \right), \\
S_{2} & = &   \alpha_c \rho_p \rho_n \left( \vec{ V}_n  - \vec{ V}_p \right) - \rho_p \left( \alpha_r \rho_p \vec{ V}_p  - \alpha_i \rho_n \vec{ V}_n \right) ,  \\
S_{3} & = &  \alpha_c \rho_p \rho_n \left( \vec{ V}_n  - \vec{ V}_p \right) \vec{ V}_p .
\label{eqn_567}
\end{eqnarray}
Next we present the equations for the ion fluid
\begin{eqnarray}
\frac{\partial \rho _p }{\partial t} + \nabla  \cdot \left( {\rho _p \vec{ V}_p } \right) & = & S_{1}, \\
\frac{\partial \left( \rho_p \vec{ V}_p \right) }{\partial t} + \nabla \cdot \left(\rho_p \vec{ V}_p \vec{ V}_p \right) - \vec{ J} \times \vec{ B} + \nabla P_p & = & S_{2}, \\
\frac{\partial \vec{ B}}{\partial t} - \nabla  \times \left( \vec{ V}_p \times \vec{ B} \right) & = &  \eta \nabla^2 \vec{B} , \\
\frac{\partial \varepsilon_p}{\partial t} + \nabla  \cdot \left[ \left( \varepsilon_p + P_p \right)\vec{ V}_p \right]  - q_p & = &  S_{3},
\label{eqn_891011}
\end{eqnarray}
where the plasma energy is given by
\begin{equation}
\varepsilon_p = \frac{P_p}{\gamma - 1} + \frac{\rho_p |\vec{V}_p|^2}{2} + \frac{|\vec{B}|^2}{2\mu_0} ,
\label{eqn_12}
\end{equation}
the lorentz force by $\vec{J} = (\nabla \times \vec{B})/ \mu_0$ and $\eta$ is the magnetic diffusivity. The ion heat flux, which includes thermal conduction and Joule heating, is given by $q_p = \nabla ^2 ( \lambda P_p / \rho _p ) + \mu_0 \eta |\vec{J}|^2$.

\section{Numerical Scheme}

We use a newly developed two-fluid numerical code to simulate the three-dimensional dynamics of the ion and neutral fluids present in the solar photosphere and chromosphere. In this numerical code, which we refer to as \emph{TwoYama}, the fluids are coupled predominantly through ion-neutral collisions but also through ionization/recombination effects. TwoYama was developed using the previously proposed \emph{Artificial Wind} (AW) numerical scheme \citep{Sokolov_et_al_1999,Sokolov_et_al_2002}, which was successfully used by \citet{Sakai_et_al_2006} to simulate the magnetic reconnection of coalescing chromospheric current loops. The AW scheme is based on the fundamental physical invariance, Galilean (or more generally Lorentz), of the governing plasma equations. It works by choosing the frame of reference such that the flow under consideration is always supersonic, thus trivializing the upwinding process and enabling highly simplified forms of shock-capturing numerical schemes to be used. In practise this is achieved by adding an artificial velocity (hence Artificial Wind) to the flow under consideration. The result is a highly accurate shock-capturing numerical code, that can simulate chromospheric problems substantially faster than other well known shock-capturing codes.

The TwoYama code was designed specifically to make use of MPI (Message Passing Interface) parallelization, so that large 3D simulations could be quickly simulated. The ionization/recombination source terms seen in the previous section, are solved using a time-splitting method at each half-time step, whilst the Joule heating, heat conduction, and magnetic diffusion terms are solved explicitly inside the AW scheme. To ensure unconditional stability of the stiff ion-neutral drag terms, we use a time-implicit differencing method, adapted from \citet{Stone_1997} and \citet{Toth_1995} to include ionization/recombination. The update of the ion/neutral momentum after each timestep, $\Delta t$, is given by
\begin{eqnarray}
\vec{R} & = &\frac{\rho _p \left( \alpha_i \Delta t\rho_n \vec{V}_n  - \alpha_r \Delta t\rho_p \vec{V}_p \right) + \alpha_c \Delta t \left( \rho_p \rho_n \vec{V}_n - \rho_n \rho_p \vec{V}_p \right)}{1 + \rho _p \left( \alpha_r \Delta t + \alpha_i \Delta t \right) + \alpha_c \Delta t \left( \rho_n  + \rho_p \right)}, \nonumber
\end{eqnarray}
\begin{eqnarray}
\rho_p^{n + 1} \vec{V}_p^{n + 1} = \rho_p \vec{V}_p + \vec{R} , & & \quad \rho_n^{n + 1} \vec{V}_n^{n + 1} = \rho_n \vec{V}_n - \vec{ R}.
\label{eqn_13}
\end{eqnarray}
Following \citet{Toth_1994}, we also use the implicitly solved momentum change, $\vec{ R}$, to update the collisional heating source terms in the ion/neutral energy equations
\begin{eqnarray}
e_p^{n + 1} = e_p^n + \vec{V}_p \cdot \vec{R}, & & \quad e_n^{n + 1} = e_n^n - \vec{V}_n \cdot \vec{ R}.
\label{eqn_14}
\end{eqnarray}
Continuous (zero-gradient) boundary conditions are used for the simulation boundaries. TwoYama solves the normalized forms of the equations seen in the previous section, according to the normalization given in the table below ($m_p$ is the proton mass).
\begin{table}[h]
\centering
\caption{Normalization Constants}
\begin{tabular}{@{} c c c l @{}}
\hline\hline
Symbol & Quantity & Constant & Value \\
\hline
$t$      & Time            & $\tau_0$                       & $=1$~s                                     \\
$n$      & Number density  & $n_0$                          & $=5\times10^{16}$~m$^{-3}$                 \\
$T$      & Temperature     & $T_0$                          & $=8400$~K                                  \\
$\rho$   & Density         & $\rho_0 = n_0 m_p$             & $\approx 8.4\times10^{-11}$~kg~m$^{-3}$    \\
$P$      & Pressure        & $P_0 = n_0 k_B T_0$            & $\approx 5.8\times10^{-3}$~Pa              \\
$V$      & Velocity        & $c_{0} = (P_0/\rho_0)^{1/2}$   & $\approx 8.3$~km~s$^{-1}$                  \\
$x,y,z$  & Length          & $\Delta = c_{0} \tau_0$        & $\approx 8.3$~km                           \\
$B$      & Magnetic field  & $B_0 = (2 \mu_0 P_0)^{1/2}$    & $\approx 1.2\times10^{-4}$~T               \\
\hline
\end{tabular}
\end{table}

\section{Simulation Model}

In this study, we assume a plasma temperature of $T_p = T_n = 8400$~K and a plasma number density of $n_p = n_n = 5 \times 10^{16}$~m$^{-3}$. Using these parameters, the background sound speed $c_s = (\gamma P_p/\rho_p)^{1/2} \approx 10.8$~km~s$^{-1}$. For simplicity, we choose a normalization time of $\tau_0 = 1$~s, such that the grid spacing $\Delta \approx 10.8$~km. Therefore the system resolution of $N_x = N_y = 1024$ corresponds to a system size of $L_x = \Delta N_x = L_y = \Delta N_y \approx 11.0$~Mm. The initial background magnetic field, pressure and density are given respectively by $B_{00}=1.0\times10^{-4}$~T, $P_{00}= n_{p} k_B T_p \approx 5.8\times10^{-3}$~Pa and $\rho_{00} = n_p m_p \approx 8.4\times10^{-11}$~kg~m$^{-3}$. Therefore, the corresponding background Alfv\'en and fast-magnetosonic speeds are respectively $c_a = B_0/(\mu_0 \rho_p)^{1/2} \approx 9.8$~km~s$^{-1}$ and $c_m = ({c_{s}}^2+{c_{a}}^2)^{1/2} \approx 14.5$~km~s$^{-1}$.

We simulate the 2.5D coalescence of two current loops in the upper chromosphere. The two current loops ($i = 1, 2$), each of radius $a = 140 \Delta \approx 1.5$~Mm, are located parallel to the z-axis and are each assumed to be initially in independent equilibrium states ($\nabla P_i=\vec{ J}_i\times \vec{ B}_i$). The initial magnetic field components of each loop (see Fig.~1) are given by
\begin{eqnarray}
B_{x} = q_i B_{z} \left(y-y_{ci}\right)/a, && B_{y} = - q_i B_{z} \left(x-x_{ci}\right)/a, \nonumber
\end{eqnarray}
\begin{equation}
B_{z} = h_i B_{00} e^{-(r_i/a)^2},
\label{eqn_15}
\end{equation}
where $r_i = [(x-x_{ci})^2+(y-y_{ci})^2]^{1/2}$, and the centers of the two current loops are $(x_{c1},y_{c1})=(350\Delta, 512\Delta)$ and $(x_{c2},y_{c2})=(674\Delta, 512\Delta)$. The initial ion pressure of each loop is given by
\begin{equation}
P_{pi} = P_{00} + P_{00}\left( {\frac{q_i^2}{2}}-{\frac{q_i^2 r_i^2}{a^2}}-1 \right) e^{-2 (r_i/a)^2},
\label{eqn_16}
\end{equation}
while the initial ion density is similarly given by
\begin{equation}
\rho_{pi} = \rho_{00} + \rho_{00}\left( {\frac{q_i^2}{2}}-{\frac{q_i^2 r_i^2}{a^2}}-1 \right) e^{-2 (r_i/a)^2}.
\label{eqn_17}
\end{equation}
The twist parameter, $q_i$, is fixed at $q_1 = q_2 = 1.0$. The helicity parameter, $h_i = \pm 1.0$, determines the orientation of the magnetic field's z-component, $B_z$. This parameter is important, as two types of magnetic reconnection can occur during the coalescence of two current loops; complete magnetic reconnection (which we call the counter-helicity case) where the $B_z$ components are anti-parallel, and partial magnetic reconnection (which we call the co-helicity case) where the $B_z$ components are parallel. The initial neutral density and pressure are set isotropically to a multiple of the background density, $\rho_{00}$, according the specific neutral-hydrogen to proton density ratio used in each simulation. The initial velocity of both the ion and neutral fluids is zero.

The heat transfer constant, seen in the heat flux terms of Sect.~2 is held fixed in all simulations at $\lambda = 1.5 \times 10^{-6}$~kg~m$^{-1}$~s$^{-1}$. The thermal conductivity coefficient can be obtained from the heat transfer constant from $\kappa = k_B \lambda / m_p \approx 0.012$~W~m$^{-1}$~K$^{-1}$. The ion-neutral collision frequency is given by $\nu_{pn} = n_p \sigma_{pn} v_{tp}$, where the collision cross-section, $\sigma_{pn} \approx 5.5 \times 10^{-19}$~m$^2$, and the ion thermal velocity, $v_{tp} \approx c_s = 10.8$~km~s$^{-1}$. Therefore $\nu_{pn} \approx 300$~Hz, giving a normalized ion-neutral collision coefficient $\tilde{\alpha}_c = \alpha_c \rho_0 \tau_0 = 300$, where $\alpha_c = \nu_{pn}/\rho_p$. If we assume that radiative recombination is dominant, then using the formula by \citet{Kaplan_Pikelner_1970}; $\alpha_r = 1.6\times 10^{8} (10^4/T)^{0.85}$~m$^{3}$~kg$^{-1}$~s$^{-1}$, which gives a normalized recombination coefficient of $\tilde{\alpha}_r = \alpha_r \rho_0 \tau_0 \approx 0.015$. Initially, we assume that recombination balances with ionization, $\tilde{\alpha}_r = \tilde {\alpha}_i = \alpha_i \rho_0 \tau_0$ and that $\rho_p = \rho_n$.

Reconnection of magnetic fields occurs due to a non-zero magnetic diffusivity (or a finite conductivity). If we take the photospheric electron-hydrogen collision frequency as $\nu_{en} \approx 10^{9}-10^{10}$~Hz \citep{Vranjes_et_al_2008}, then we can estimate the magnetic diffusivity due to collisions between electrons and hydrogen atoms as
\begin{equation}
\eta = \frac{\nu_{en} m_e}{\mu_0 n_e e^2 } \approx 5.6 \times 10^{5-6} \quad \left(\mathrm{m}^2 \mathrm{s}^{-1}\right).
\label{eqn_18}
\end{equation}
If we take the characteristic length scale as $\Delta \approx 10.8$~km, and the characteristic velocity as the sound speed $c_s \approx 10.8$~km~s$^{-1}$, then the magnetic diffusivity corresponds to a magnetic Reynolds number $R_m = \Delta c_s / \eta \approx 20-200$. In this study $\eta = 1.0 \times 10^7$~m$^2$~s$^{-1}$, which corresponds to $R_m \approx 10$.

To gauge the rate of magnetic reconnection during the coalescence of the two current loops, we measure the total reconnected magnetic flux, $\psi$, at the point $x = y = 512\Delta$, using the following expression
\begin{equation}
\psi = - \int{E_z}dt,
\label{eqn_19}
\end{equation}
where $E_z$ is the induced electric field.

\section{Simulation Results}
In this section we present the results of our numerical simulations. Unless stated, all of the results are given in normalized units, according the normalization specified in Table~1.

\begin{figure}[t]
\includegraphics[width=\linewidth]{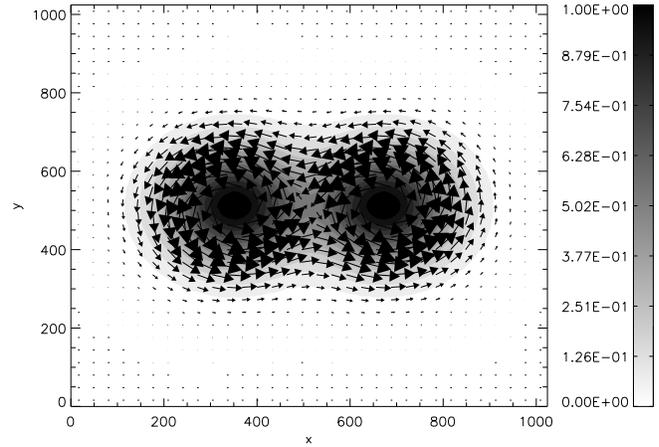}
\caption{The initial magnetic field: a contour plot of $B_z$ overlayed by a vector ($B_x$~--~$B_y$) plot.}
\label{initbfield}
\end{figure}

\begin{figure}[t]
\includegraphics[width=\linewidth]{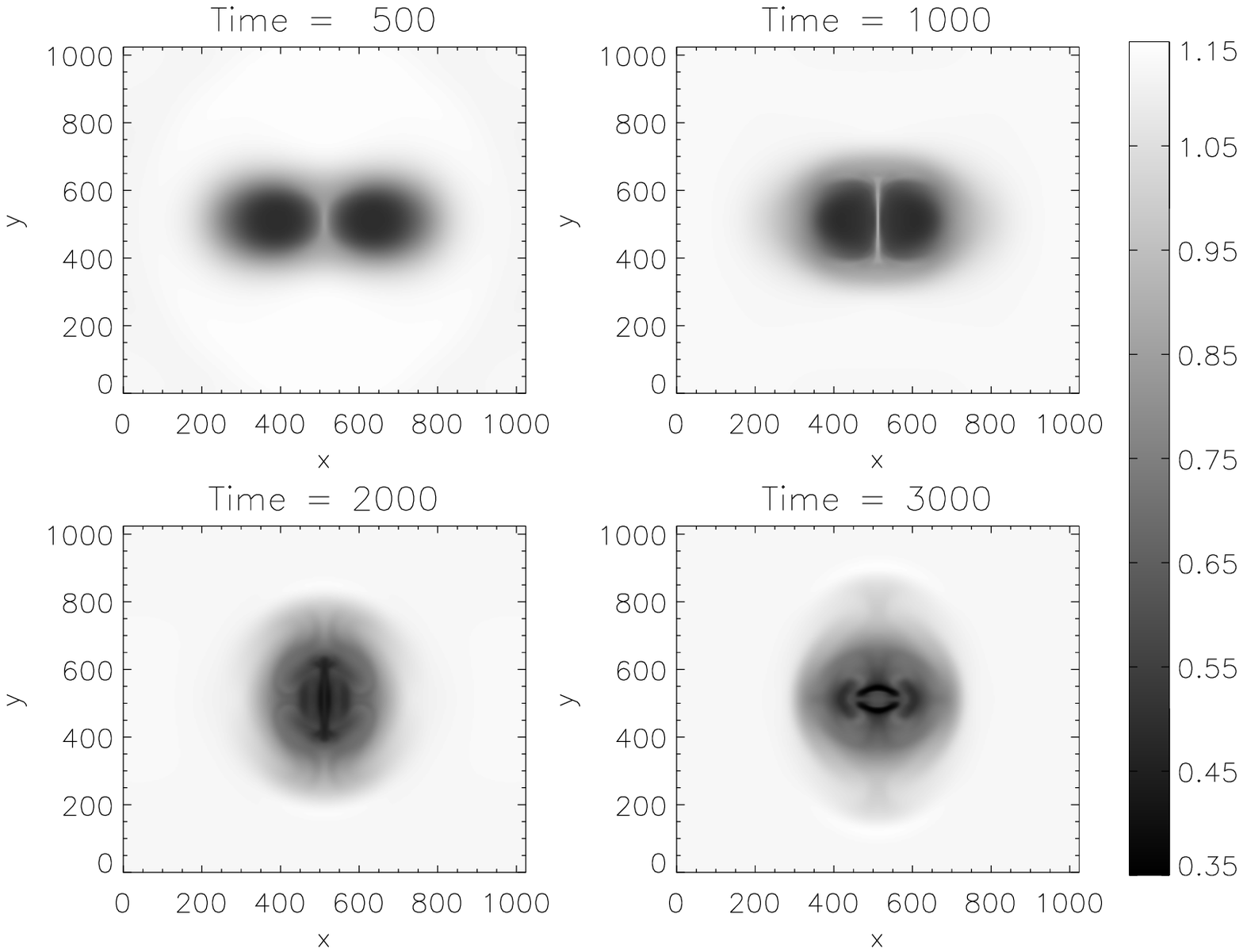}
\caption{Time development of the ion pressure for two co-helical current loops, where $\rho_n/\rho_p = 1$.}
\label{cnt_ionp}
\end{figure}

\begin{figure}[t]
\includegraphics[width=\linewidth]{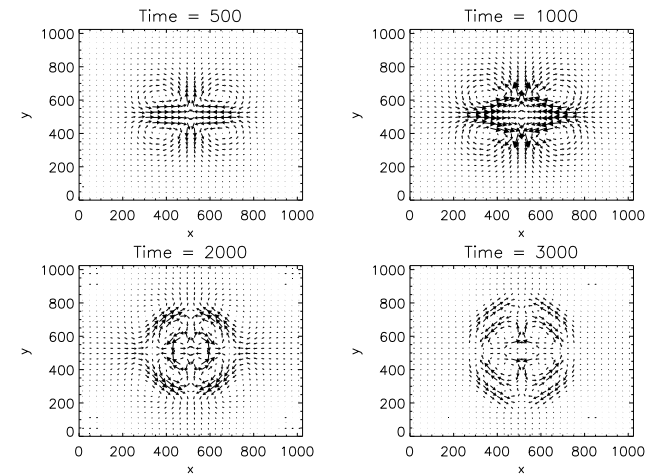}
\caption{Time development of the ion velocity ($V_x$~--~$V_y$) for two co-helical current loops, where $\rho_n/\rho_p = 1$.}
\label{vec_v}
\end{figure}

\begin{figure}[t]
\includegraphics[width=\linewidth]{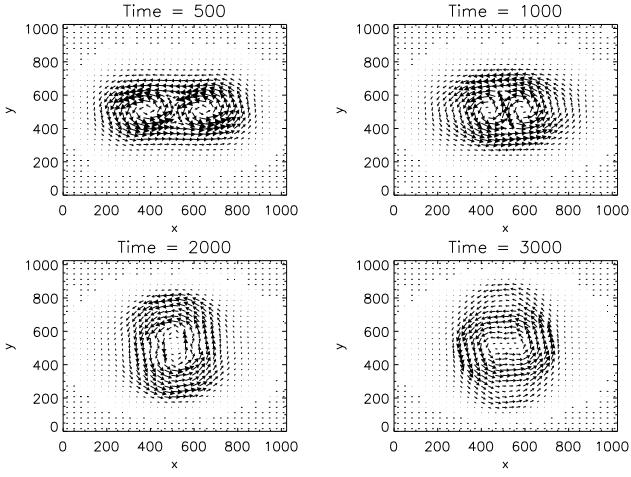}
\caption{Time development of the magnetic field ($B_x$~--~$B_y$) for two co-helical current loops, where $\rho_n/\rho_p = 1$.}
\label{vec_b}
\end{figure}

\begin{figure}[t]
\includegraphics[width=\linewidth]{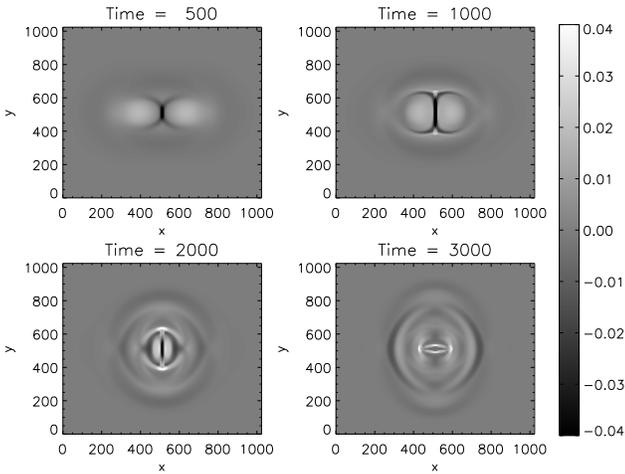}
\caption{Time development of the current density, $J_z$, for two co-helical current loops, where $\rho_n/\rho_p = 1$.}
\label{cnt_jz}
\end{figure}

\begin{figure}[t]
\includegraphics[width=\linewidth]{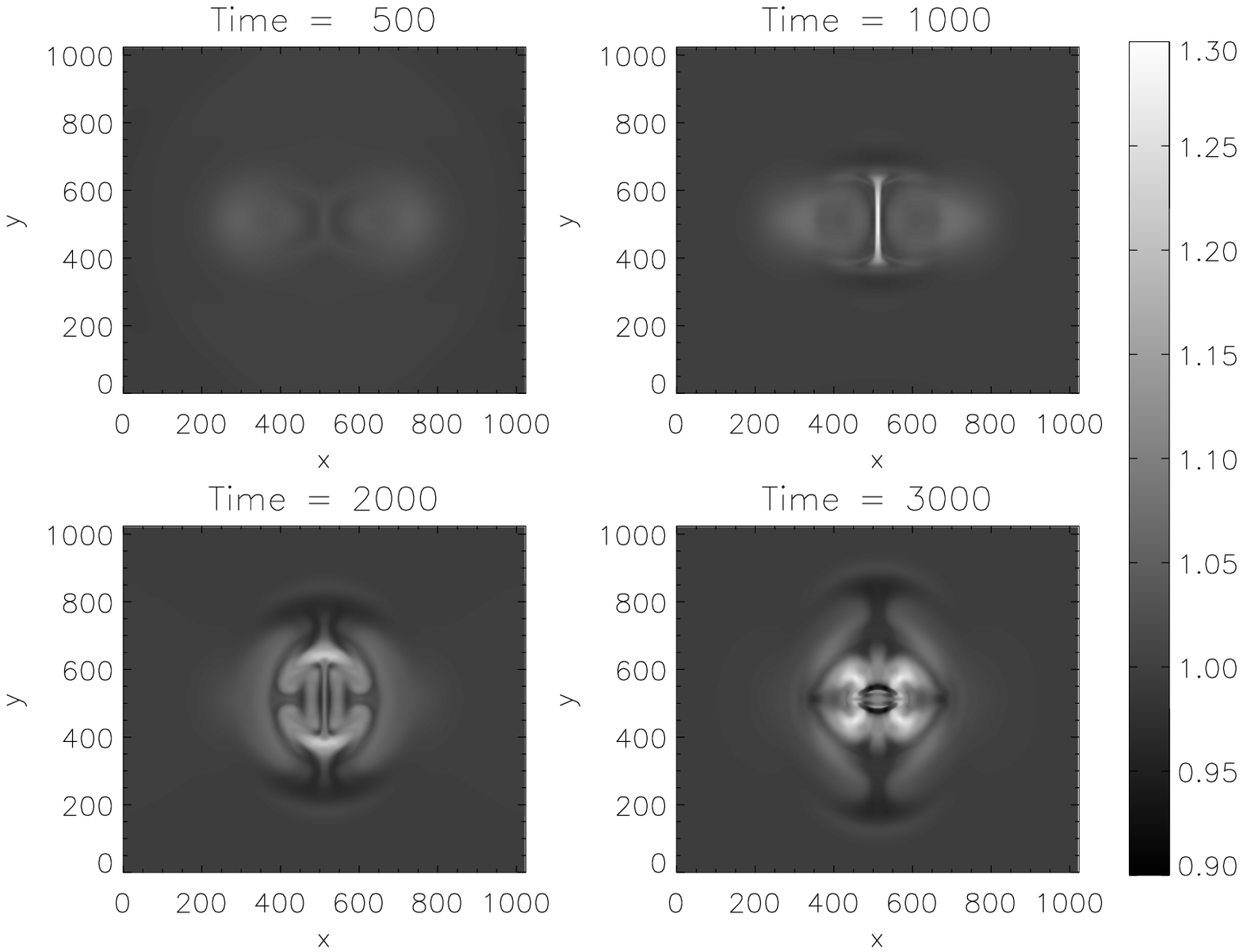}
\caption{Time development of the ion temperature for two co-helical current loops, where $\rho_n/\rho_p = 1$.}
\label{cnt_iont}
\end{figure}

\begin{figure}[t]
\includegraphics[width=\linewidth]{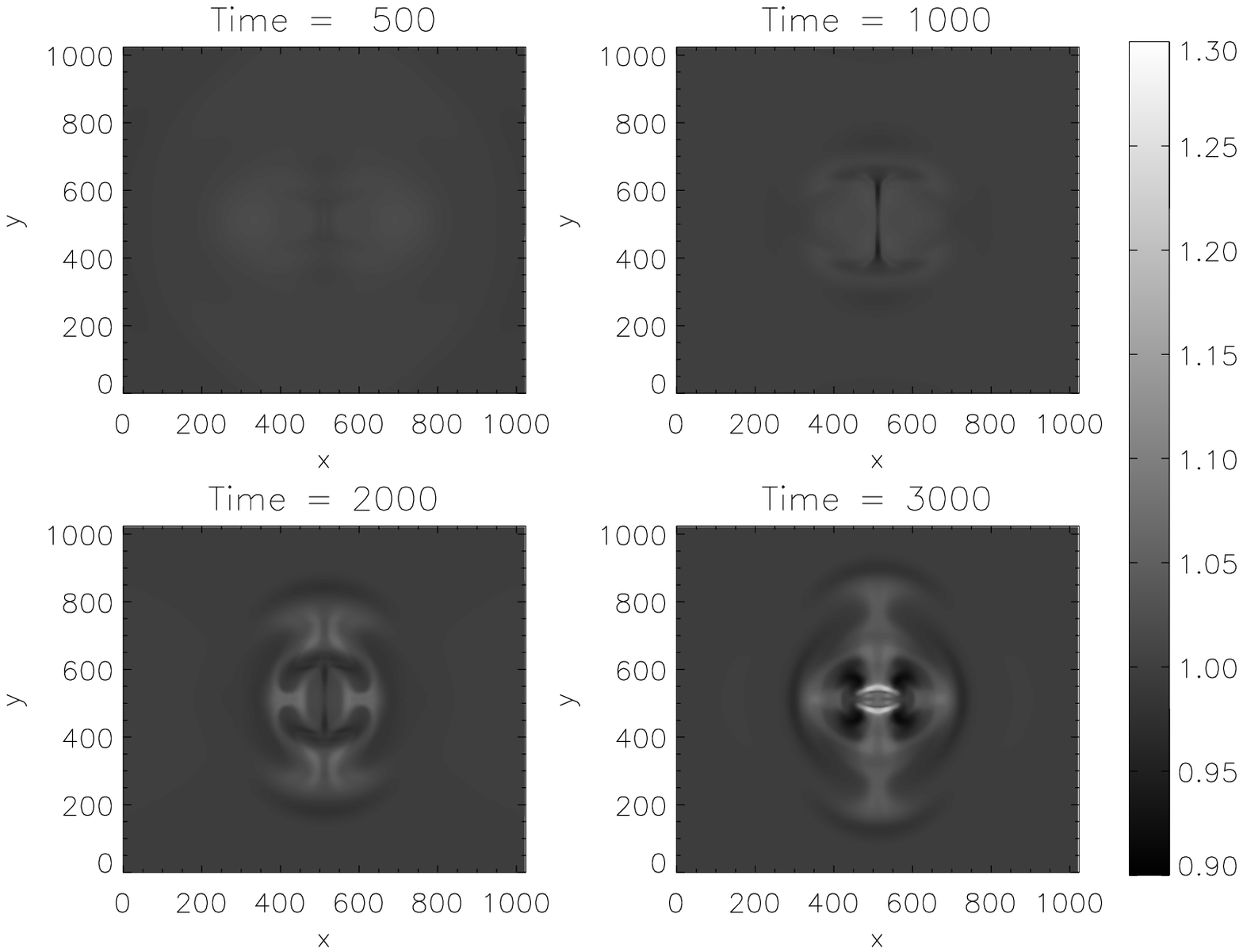}
\caption{Time development of the neutral temperature for two co-helical current loops, where $\rho_n/\rho_p = 1$.}
\label{cnt_ntlt}
\end{figure}

\begin{figure}[t]
\includegraphics[width=\linewidth]{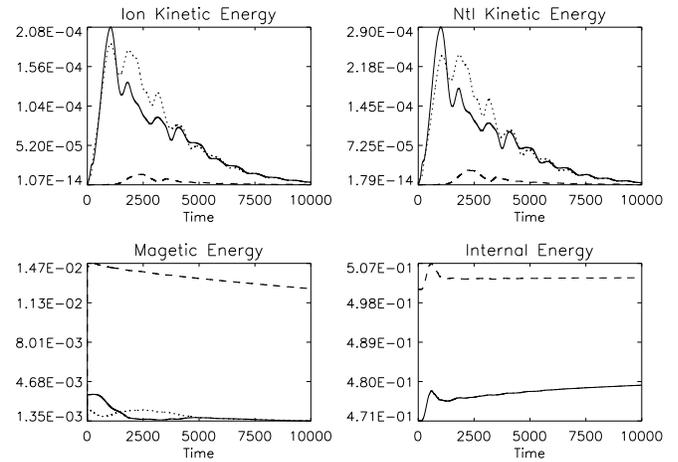}
\caption{Time development of plasma energy for two co-helical current loops ($\rho_n/\rho_p = 1$). Kinetic and magnetic energy plots: (solid) x-component, (dotted) y-component and (dashed) z-component. Internal energy plot: (solid) ions and (dashed) neutral. Each energy component is normalized to the total plasma energy.}
\label{energy}
\end{figure}

We begin with Fig.~\ref{initbfield}, where we show the initial magnetic field configuration for two co-helical current loops. The initial ion pressure in-balance between the two loops (Fig.~\ref{cnt_ionp}), causes them to spontaneously collapse towards each other. During the initial coalescence phase ($t\le1000$), strong inflows (Fig.~\ref{vec_v}) develop along the x-axis, with subsequent outflows along the y-axis. This inflow of plasma compresses oppositely directed magnetic field lines, seen in the central region of Fig.~\ref{vec_b}, to form a strong reverse current sheet (see Fig.~\ref{cnt_jz}) parallel to the y-axis. The current sheet is then dissipated by the plasma's finite conductivity, leading to reconnection of the ($B_x-B_y$) magnetic field lines. The significant heating (Joule and collisional) generated during the reconnection process, is clearly visible in Fig.~\ref{cnt_iont}, where we see a 30\% rise in ion temperature along the current sheet. Conversely, in Fig.~\ref{cnt_ntlt} we see a 5-10\% reduction in neutral temperature along the current sheet, due to collisional cooling. This implies that 20-25\% of the increase in ion temperature is due to Joule heating, while a further 5-10\% increase is due to collisional heating with neutral particles. In Fig.~\ref{energy} we see that during the initial coalescence phase, reconnection of magnetic fields lines reduces the magnetic energy, or magnetic pressure, relative to the ion thermal pressure. This is turn accelerates the ion and neutral fluids (due to ion-neutral collisions) into the central region; an effect clearly seen in Fig.~\ref{energy} by the rapid increase in ion and neutral kinetic energy. In the later coalescence stage ($t>1000$), collisional heating is seen to slowly convert the ion and neutral kinetic energy into internal (thermal) plasma energy. Fig.~\ref{energy} demonstrates that the main effect of the magnetic reconnection process is the conversion of magnetic energy into kinetic and thermal energy.

\subsection{Effects of ion-neutral density ratio}
Firstly, we investigate how the rate of magnetic reconnection is altered by increasing the neutral-hydrogen to proton density ratio, $\rho_n/\rho_p$. We present numerical results for the coalescence of two co-helical current loops, using density ratios of $\rho_n/\rho_p = 1, 10, 100, 1000$. The normalized collision frequency is held fixed at $\tilde{\alpha}_c = 300$, while ionization and recombination effects are ignored ($\tilde{\alpha}_i = \tilde{\alpha}_r = 0.0$).

\begin{figure}[t]
\includegraphics[width=\linewidth]{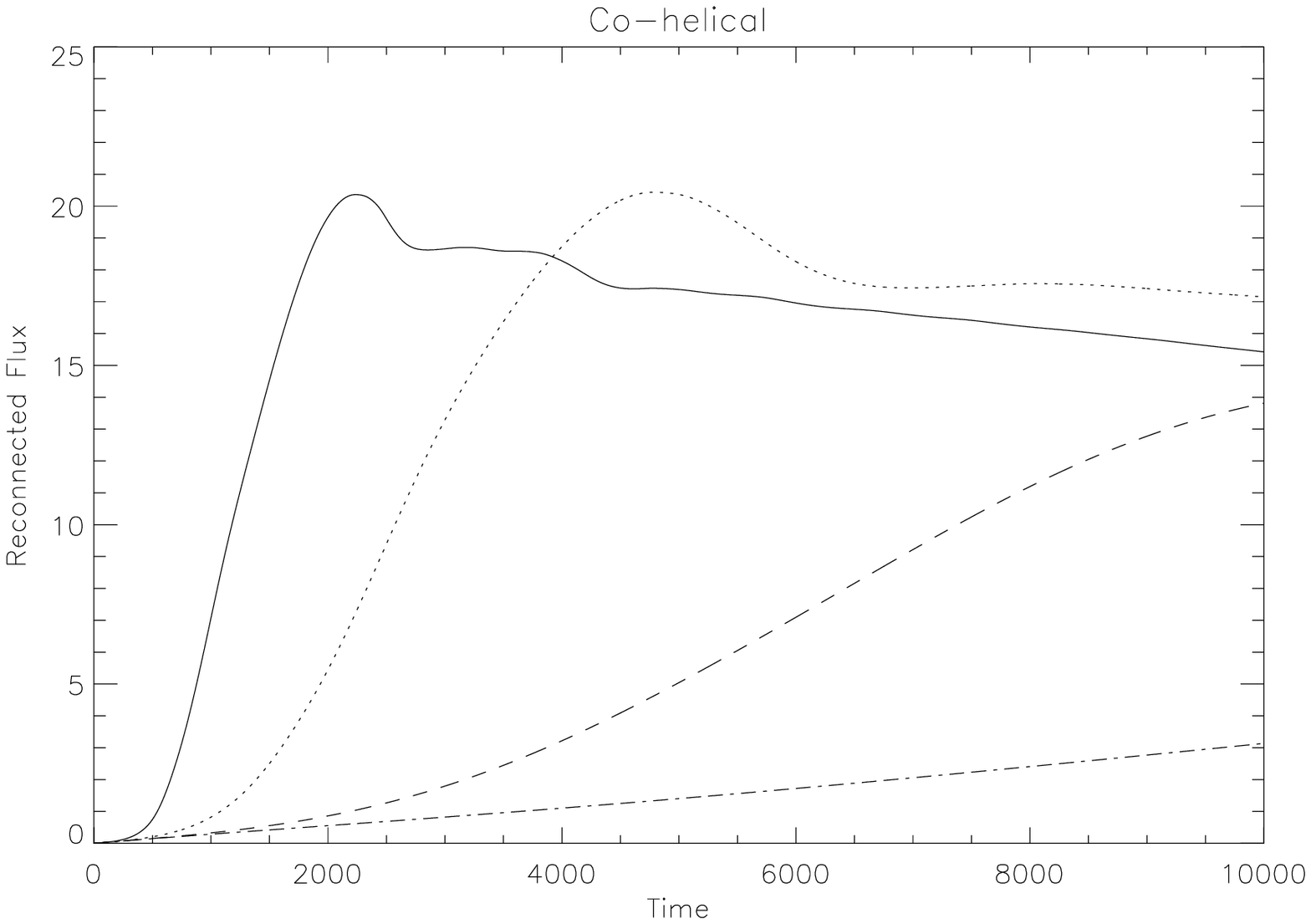}
\caption{Effect of neutral-hydrogen to proton density ratio on the reconnected magnetic flux for two co-helical current loops. (solid) $\rho_n/\rho_p = 1$, (dotted) $\rho_n/\rho_p = 10$,  (dashed) $\rho_n/\rho_p = 100$ and (dot-dashed) $\rho_n/\rho_p = 1000$.}
\label{recflux1}
\end{figure}

It is well known that the neutral-hydrogen to proton density ratio increases with depth in the chromosphere/photosphere. Changing this density ratio therefore effectively corresponds to a change in depth at which the simulated current loop coalescence, and subsequent magnetic reconnection takes place. The density ratio $\rho_n/\rho_p = 1$ is representative of the upper chromosphere, while $\rho_n/\rho_p = 1000$ is representative of the lower chromosphere. In Fig.~\ref{recflux1}, we use a normalized form of Eq.~19 to plot the total reconnected magnetic flux as a function of time, for various neutral-hydrogen to proton density ratios. The slope of each line is a direct measure of the magnetic reconnection rate (as measured at $x=y=512\Delta$). From this figure, it is immediately clear that increasing the neutral-hydrogen to proton density ratio strongly decreases the magnetic reconnection rate. Comparing the lines representing $\rho_n/\rho_p = 1$ and $\rho_n/\rho_p = 10$, we see that a ten-fold increase in density ratio reduces the magnetic reconnection rate by a factor of two. Whilst, comparing the lines representing $\rho_n/\rho_p = 1$ and $\rho_n/\rho_p = 1000$, we see that a thousand-fold increase in density ratio reduces the magnetic reconnection rate by a factor of twenty. The simulation results seen in Fig.~\ref{recflux1} therefore suggest that, in the case of coalescing current loops, magnetic reconnection proceeds faster in the upper chromosphere than in the lower chromosphere.

Fig.~\ref{recflux1} shows a direct correlation between magnetic reconnection rate and density ratio, however the correlation between the peak reconnected magnetic flux, $\psi$, and the density ratio is less clear. The peak value for the lines $\rho_n/\rho_p = 1$ and $\rho_n/\rho_p = 10$ are approximately equivalent at $\psi \approx 21$, whereas the lines representing the higher density ratios of $\rho_n/\rho_p = 100 - 1000$, appear to indicate a reduction in the peak reconnected magnetic flux with increasing density ratio.

Finally, comparisons of magnetic and velocity vector plots for various density ratios, in addition to Fig.~\ref{recflux1}, appear to show that the density of neutral particles does not alter the fundamental dynamics of coalescence process itself, but simply the rate at which it occurs.

\subsection{Effects of ionization/recombination}
The recombination of protons and electrons to form neutral hydrogen, is an important physical process occurring in the partially ionized upper chromosphere. We therefore investigate how the magnetic reconnection rate is altered by including the effects of ionization and recombination. We show results for two co-helical current loops using a density ratio of $\rho_n/\rho_p = 1$. The normalized collision frequency is fixed at $\tilde{\alpha}_c = 300$, while the normalized ionization and recombination coefficients are set to $\tilde{\alpha}_i = \tilde{\alpha}_r = 1.5, 0.15, 0.015, 0.0$.

\begin{figure}[t]
\includegraphics[width=\linewidth]{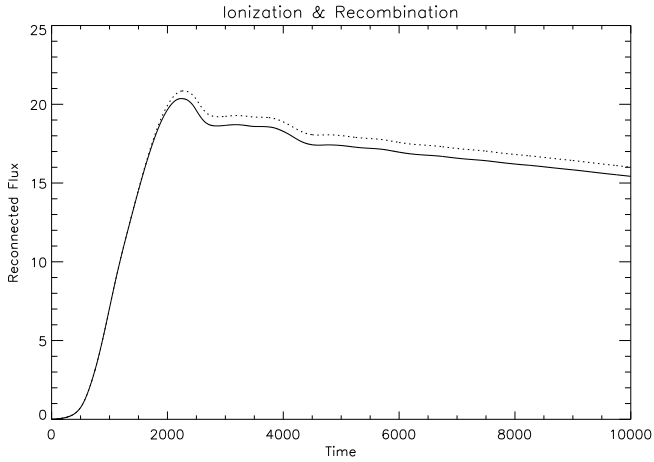}
\caption{Effect of ioniziation/recombination on the the reconnected magnetic flux for two co-helical current loops ($\rho_n/\rho_p = 1$). (solid) $\tilde{\alpha}_i = \tilde{\alpha}_r = 0.0$ and (dotted) $\tilde{\alpha}_i = \tilde{\alpha}_r = 0.015$.}
\label{recflux2}
\end{figure}

In Fig.~\ref{recflux2}, for clarity we do not show the two cases of $\tilde{\alpha}_i = \tilde{\alpha}_r = 0.15, 1.5$ since these gave visually indistinguishable results to the $\tilde{\alpha}_i = \tilde{\alpha}_r = 0.015$ case. From the figure we see that the inclusion of ionization/recombination effects has no effect on the rate of magnetic reconnection. We do however see an increase in the total reconnected magnetic flux by $\approx 5\%$. This phenomena was previously explained by \citet{Sakai_1996} as resulting from a decrease in ion pressure at the current sheet, formed during the coalescence process (see Fig.~\ref{cnt_jz}). This decrease in ion pressure results from protons and electrons in the current sheet recombining to form neutral hydrogen. Since it is the magnetic field which drives the reconnection process, any reduction of the ion pressure, which alone balances the magnetic pressure, will result in enhanced magnetic reconnection.

\subsection{Effects of collision frequency}
Next we investigate how the magnetic reconnection rate is altered by changing the ion-neutral collision frequency. We show simulation results for two co-helical current loops using normalized collision frequencies of $\tilde{\alpha}_c = 300, 3.0, 0.03, 0.0$. The density ratio is fixed at $\rho_n/\rho_p = 1$, while ionization and recombination effects are ignored ($\tilde{\alpha}_i = \tilde{\alpha}_r = 0.0$).

\begin{figure}[t]
\includegraphics[width=\linewidth]{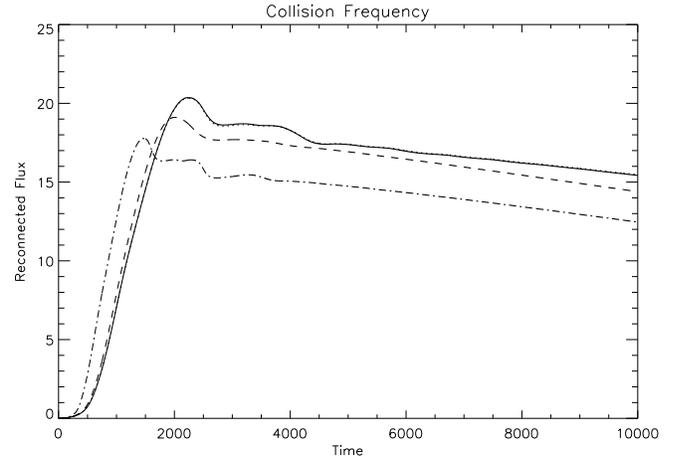}
\caption{Effect of ion-neutral collision frequency on the reconnected magnetic flux for two co-helical current loops ($\rho_n/\rho_p = 1$). (solid) $\tilde{\alpha}_c = 300$, (dotted) $\tilde{\alpha}_c = 3.0$, (dashed) $\tilde{\alpha}_c = 0.03$ and (dot-dashed) $\tilde{\alpha}_c = 0.0$ (MHD).}
\label{recflux3}
\end{figure}

Firstly, in Fig.~\ref{recflux3}, we note that the lines representing $\tilde{\alpha}_c = 300$ and $\tilde{\alpha}_c = 3$ produce nearly identical results (only $\tilde{\alpha}_c = 300$ is visible). This is because both cases represent strongly coupled ion-neutral plasmas; the ion-neutral collisions cause the ion and neutral particles to behave as a single \emph{heavy} fluid with a combined ion-neutral density. For the $\tilde{\alpha}_c = 0.0$ case on the other hand, the ion and neutral particle dynamics are completely decoupled. The ions therefore move as a single \emph{light} fluid (equivalent to a MHD simulation). The collapse and eventual coalescence of the two current loops is driven by a continuous reduction in magnetic pressure (due to magnetic reconnection), balanced only by the ion thermal pressure. The neutral particles by definition are not influenced by the magnetic field, and therefore play no part in balancing the magnetic pressure. However, the effective inertia of the initially stationary fluids, does depend on both the neutral-hydrogen to proton density ratio and the ion-neutral collision frequency. Given that the initial conditions for the ion particles and magnetic field are the same in each simulation, the initial acceleration of the fluids can only therefore depend on their effective inertia. In strongly coupled ion-neutral plasmas, where the effective inertia is relatively high, we would expect the coalescence and subsequent magnetic reconnection to proceed, at least initially, relatively slowly. Whereas in weakly coupled ion-neutral plasmas, where the effective inertia is relatively low, we would expect the reverse to be true. This appears to be confirmed by Fig.~\ref{recflux3}, where we see a slight increase in the rate of magnetic reconnection, with decreasing ion-neutral collision frequency.

In Fig.~\ref{recflux3}, the rate of magnetic reconnection for the MHD case ($\tilde{\alpha}_c = 0.0$), does not differ significantly from the two-fluid cases ($\tilde{\alpha}_c = 300, 3.0, 0.03$). We note however, that this is only true for the low neutral density case, where $\rho_n/\rho_p = 1$. If we compare the MHD case of Fig.~\ref{recflux3} to the $\rho_n/\rho_p = 1000$ case of Fig.~\ref{recflux1}, we see a twenty-fold difference in the magnetic reconnection rates. The MHD approximation is therefore not an appropriate model to describe magnetic reconnection in the lower chromosphere or photosphere.

Finally in Fig.~\ref{recflux3}, we note that the peak reconnected magnetic flux appears to decrease with decreasing ion-neutral collision frequency. We speculate that the increased effective inertia of ion-neutral fluids may extend the time for loop coalescence, thus leading to an increased reconnected magnetic flux. This effect is of course balanced by the reduced initial acceleration, caused by increased neutral density, discussed above. The results of Fig.~\ref{recflux1} would therefore suggest that neutral-hydrogen to proton density ratios of $\rho_n/\rho_p = 1 - 10$, would lead to optimum amounts of reconnected magnetic flux.

\subsection{Effects of loop helicity}
In this final subsection, we investigate how the magnetic reconnection rate is altered by changing the relative helicities of the two current loops. We show simulation results for two counter-helical (opposite $B_z$) current loops, using density ratios of $\rho_n/\rho_p = 1, 10, 100, 1000$. The normalized collision frequency is held fixed at $\tilde{\alpha}_c = 300$, while ionization and recombination effects are ignored ($\tilde{\alpha}_i = \tilde{\alpha}_r = 0.0$).

\begin{figure}[t]
\includegraphics[width=\linewidth]{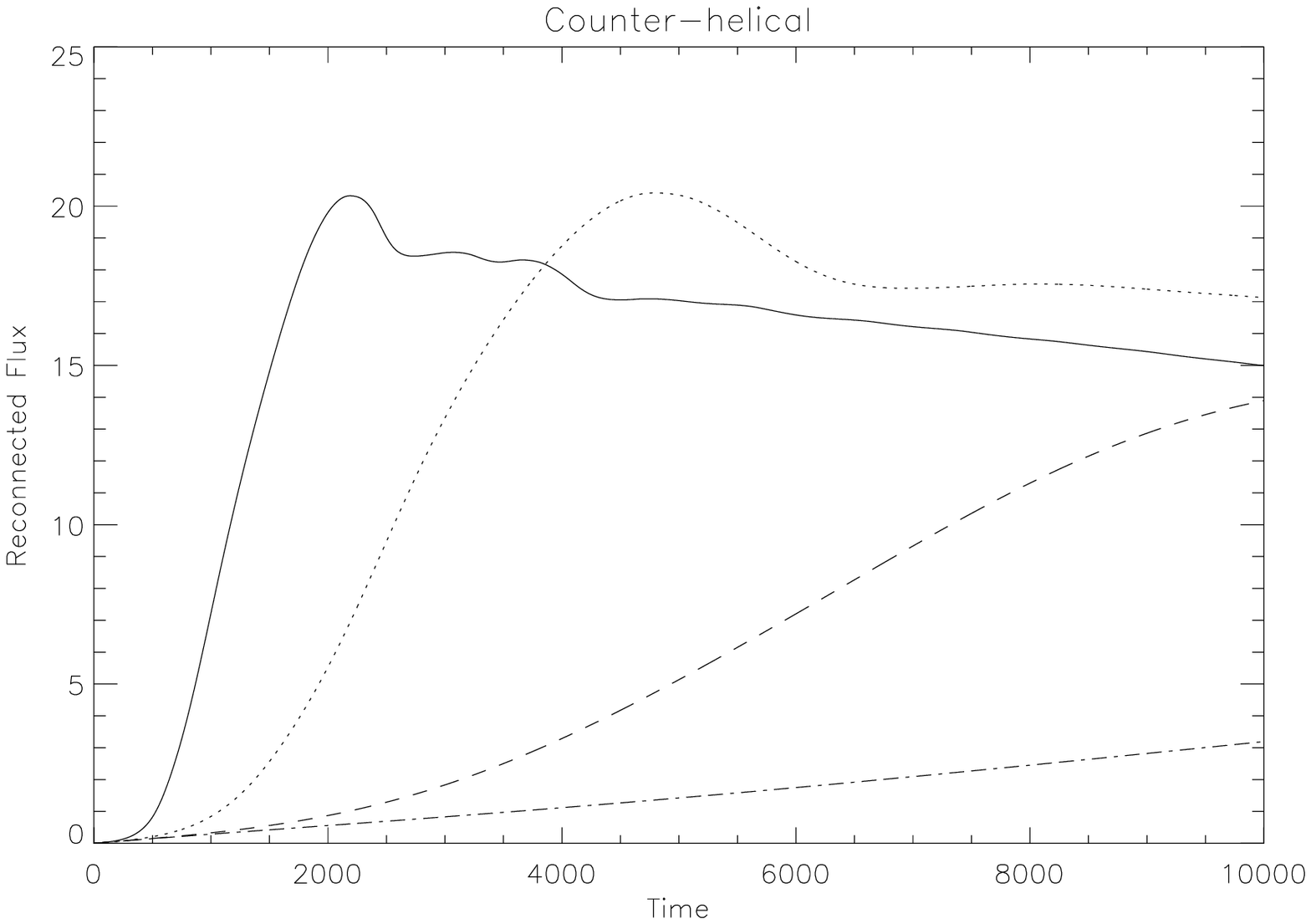}
\caption{Effect of neutral-hydrogen to proton density ratio on the reconnected magnetic flux for two counter-helical current loops. (solid) $\rho_n/\rho_p = 1$, (dotted) $\rho_n/\rho_p = 10$, (dashed) $\rho_n/\rho_p = 100$ and (dot-dashed) $\rho_n/\rho_p = 1000$.}
\label{recflux4}
\end{figure}

In Fig.~\ref{recflux4}, we see that, as previously seen in the co-helical case, the rate of magnetic reconnection strongly depends on the density ratio. Comparisons of Fig.~\ref{recflux4} to the co-helical case seen in Fig.~\ref{recflux1}, show only minor variations in the both the rates of magnetic reconnection and the total reconnected magnetic flux. We expect that this result holds true only for the 2.5D case studied here, since previous 3D studies have shown that the relative helicities of the loops does indeed affect the magnetic reconnection process. A subsequent 3D study will investigate the relative helicity further.

\section{Discussions and Conclusions}

In this study, we investigated the magnetic reconnection of two coalescing chromospheric current loops. We conducted 2.5D simulations of the coalescence using a newly developed two-fluid (ion-neutral) code, which includes the effects of ion-neutral collisions, ionization/recombination, thermal/resistive diffusivity and collisional/resistive heating.

In the results of the previous section, we found that the fundamental physical dynamics of the coalescence process was not altered by changing (a) the neutral-hydrogen to proton density ratio, (b) the ionization/recombination coefficients, (c) the ion-neutral collision frequency or (d) the relative helicities of the two loops. We did find however, that the rate of magnetic reconnection was susceptible to each parameter to a varying degree:
\begin{itemize}
\item[(a)] Increasing the neutral-hydrogen to proton density ratio strongly decreased the rate of magnetic reconnection.
\item[(b)] The inclusion of ionization and recombination effects did not alter the rate of magnetic reconnection.
\item[(c)] Decreasing the ion-neutral collision frequency slightly increased the rate of magnetic reconnection.
\item[(d)] The relative helicities of the two current loops had no discernable effects on the rate of magnetic reconnection.
\end{itemize}
Altering the neutral-hydrogen to proton density ratio, had a significantly stronger effect on the rates of magnetic reconnection, than any of the other parameters listed above. In fact, we found that the rate of magnetic reconnection for $\rho_n/\rho_p = 1$ (upper chromosphere) was twenty times faster than for $\rho_n/\rho_p = 1000$ (lower chromosphere). This important result, which may be applicable to other forms of chromospheric magnetic reconnection, implies that observed jets associated with fast magnetic reconnection, occur preferentially in the upper chromosphere / lower corona.
Comparisons of two-fluid and MHD (MagnetoHydroDynamic) simulation results, show significant differences in the measured rates of magnetic reconnection, particularly for the density ratios representative of the lower chromosphere. This demonstrates that MHD is not an appropriate model to describe magnetic reconnection in the solar chromosphere/photosphere.

The main result of this study shows that the rates of magnetic reconnection for coalescing current loops, are strongly affected by the inclusion of ion-neutral collisions. We therefore conclude that it is essential that the effects of neutral particles are included in future analytical/numerical models of chromospheric magnetic reconnection.

\begin{acknowledgements}
This research was supported by the Japan Society for the Promotion of Science (JSPS).
\end{acknowledgements}

\bibliography{Smith}

\newpage

\end{document}